# Early Prediction of AI-Assisted Cheating Risk in Online Exams Through Learning Analytics

Gökhan Akçapınar
Department of Computer Education and Instructional Technology
Hacettepe University
Ankara, Türkiye
gokhana@hacettepe.edu.tr

*Abstract:* **AI-assisted cheating has become an important threat to the security of online exams. This study examines whether the risk of AI-assisted cheating in the final exam can be predicted using students' digital traces in the learning management system (LMS) during the first eight weeks of the semester. The sample comprised 52 first-year undergraduates enrolled in a bachelor's program in Computer Education and Instructional Technology and taking an Introduction to Programming course at a public university in Türkiye. Students were labeled as low- or high-risk based on suspicious behaviors recorded in the final-exam logs, including copy, focus-loss, and right-click events. Of the 52 students, 23 (44.2%) were labeled as high-risk in a proctored, face-to-face exam. Group membership was then predicted using five features selected from 27 candidates extracted from students' digital traces. Logistic Regression, Naive Bayes, Random Forest, and Gradient Boosting algorithms were used to build the prediction models. Model performance was evaluated using leave-one-out cross-validation (LOOCV) with fold-specific preprocessing and feature selection. Logistic Regression achieved the best performance (Accuracy = 73.1%). The results indicate that LMS interaction data can provide an early signal of AI-assisted cheating risk. Course-module views, assignment submissions, and the number of days on which course videos were accessed were the most consistently selected features across the LOOCV folds. These predictions are intended to support timely academic guidance, not to establish misconduct or initiate disciplinary action.**

## I. INTRODUCTION

Recent advances in GenAI offer opportunities for education but also create risks to the integrity of educational assessment [1]. Survey-based studies using indirect questioning suggest that unauthorized GenAI use is widespread among students. A survey of 95,513 students across 20 U.S. public research universities estimated that 9% of GenAI users had submitted AI-generated content as their own work despite knowing that it might not be permitted [2]. A UK study using indirect questioning estimated that approximately 22% of undergraduates had cheated with GenAI during the 2023–2024 academic year [3]. The concern is amplified by model capability: a scoping review reported that GPT-4 passed most of the included higher-education multiple-choice examinations [4], and a blind real-world study found that 94% of fully AI-generated submissions inserted into an actual university assessment system were not identified and received higher average grades than authentic student submissions [5]. At the same time, students do not uniformly interpret all forms of AI assistance as cheating, underscoring the importance of distinguishing ordinary academic use from unauthorized use [6]. The accessibility of multimodal tools and browser extensions has further reduced the effort required to obtain unauthorized assistance during online assessments [1, 3]. Responses include assessment redesign [7], AI-driven proctoring for online exams [8], and video-based anomaly detection in examination settings [9].

Proctoring approaches monitor students' behavior during examinations [8, 9], while AI-text detection approaches examine submitted work for signs of AI-generated content [10, 11]. Evaluations of AI-text detectors have identified accuracy and reliability limitations, including reduced performance after text manipulation, that limit their suitability for high-stakes academic-integrity decisions [10, 11]. A separate line of work focuses on pedagogical responses and assessment redesign [7]. Together, these studies show that GenAI presents a serious challenge to academic integrity [1]. Although assessment redesign can be implemented before an exam, it does not necessarily identify individual students who may benefit from early support. This study examines whether early-semester LMS interactions can support individual risk prediction before the final exam, drawing on learning-analytics research on engagement and academic risk [12, 13].

Studies based specifically on students' in-exam interaction patterns associated with AI-assisted cheating remain limited. In a previous study [14], behaviors such as text selection, right-clicking, and focus loss in a proctored exam were analyzed using a clustering method, and about one-third of the students showed patterns that could be associated with AI-assisted cheating. That study showed that students at high risk of AI-assisted cheating could be meaningfully distinguished through behavioral indicators in exam logs. However, because the approach identifies risk during or after the assessment, it offers limited opportunity for preventive intervention.

In online learning environments, the interaction patterns that students show throughout the semester provide rich information about academic performance, participation, and disengagement. Systematic reviews show that LMS traces are widely used to operationalize behavioral engagement and identify students at risk of poor academic outcomes, while also cautioning that such traces primarily capture observable activity rather than cognitive or emotional engagement [12, 13]. LMSs, which are widely used in hybrid and distance education environments, record students' interactions such as assignment submissions, quiz attempts, and resource viewing with time stamps. The relationship of this rich behavioral data source with students' risk of AI-assisted cheating has not yet been sufficiently investigated. Building on the behavioral distinguishability shown in the previous study [14], this study moves one step further and investigates whether students' AI-assisted cheating risk can be predicted early through their in-semester LMS interaction data, rather than through in-exam behaviors. The study addressed the following question: Can students' LMS behaviors during the first 8 weeks of the semester meaningfully predict their risk of AI-assisted cheating in the final exam?

Previous learning-analytics research has linked early behavioral engagement to later academic outcomes [12, 13]. This study extends that work by using a behaviorally derived indicator of AI-assisted cheating risk as the outcome, thereby connecting engagement analytics with academic integrity. The model uses only the first eight weeks of interaction data, which allows supportive action before the final exam. It also addresses misconduct enabled by GenAI. The study is therefore framed as an early-warning approach to academic integrity rather than as a conventional performance-prediction model.

## II. METHOD

The dataset was obtained from 52 first-year undergraduates enrolled in a bachelor's program in Computer Education and Instructional Technology at a public university in Türkiye. The students took an Introduction to Programming course in a hybrid format. The study uses the same student cohort and final-exam interaction records as the previous study [14]. It extends that work by using early-semester learning traces to predict student-level risk labels derived from question-level exam events. All videos, assignments, quizzes, and supplementary resources were provided through Moodle LMS. The final exam was administered face to face in a computer laboratory through a web-based exam system. It contained 25 multiple-choice questions, with questions and options presented in random order. Students could not return to an earlier question, and the system advanced automatically when the time allocated to a question expired.

Before the exam, students were informed that only the exam application could remain open, that activities other than reading and answering questions would violate the exam rules, and that the system would record their interactions. All records were anonymized before analysis, and no personally identifiable information was used. The study does not label students as cheaters. The target class is a behaviorally derived indicator of AI-assisted cheating risk, not a measure of confirmed misconduct.

Fig. 1 summarizes the study workflow. Exam interaction logs were used to construct the behavioral risk labels, while Moodle and custom video-player records from the first eight weeks supplied the 27 candidate predictors. Within each LOOCV fold, preprocessing and Mann-Whitney U feature selection were fitted only on the 51-student training set before a prediction was generated for the held-out student. The 52 out-of-sample predictions per classifier were then aggregated for evaluation. This fold-specific procedure prevented the test student's data from influencing preprocessing, feature selection, or model fitting.

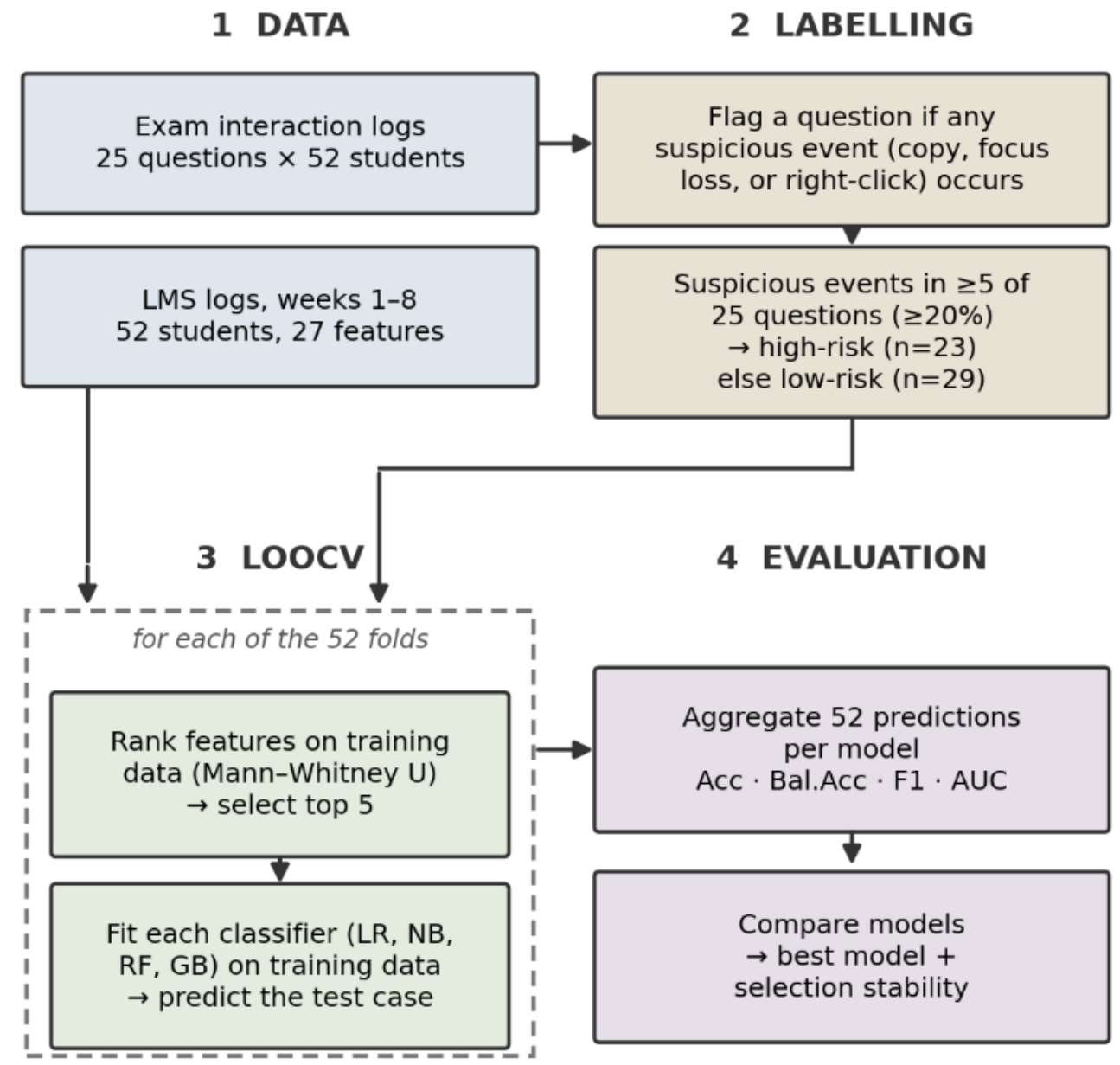

Fig. 1. Overview of the study workflow

### A. *Target Variable Construction*

In the previous study [14], aggregated suspicious-behavior indicators (the total counts of text selection, right-clicking, and focus loss across the whole exam) were analyzed with k-Means clustering, which showed that students could be meaningfully separated according to their suspicious behavior. However, because labeling was performed at the whole-exam level, it could not capture patterns that varied from question to question. In the present study, a finer-grained labeling approach was therefore adopted.

Students' AI-assisted cheating risk was used as the target variable. Three events were defined in advance as suspicious because they were prohibited by the exam instructions: copying a question or option, moving focus from the exam screen to another tab or application, and right-clicking. These events can accompany unauthorized AI use. A right-click can invoke a browser extension, a copy event can support transferring a question to an external tool, and moving to an external tool produces a focus-loss event. Text selection alone was excluded because it may occur during ordinary reading. A response was flagged as suspicious when at least one of the three events occurred while that question was displayed. The labels represent a behavioral proxy for risk rather than verified AI use.

Each question was counted at most once, regardless of how many times these interactions occurred within it. The suspicious response status was evaluated separately for each of the 25 questions in the exam, and the number of questions in which each student showed suspicious behavior across the exam was calculated. Fig. 2 presents the resulting distribution, plotting each student's suspicious question frequency against their final exam score.

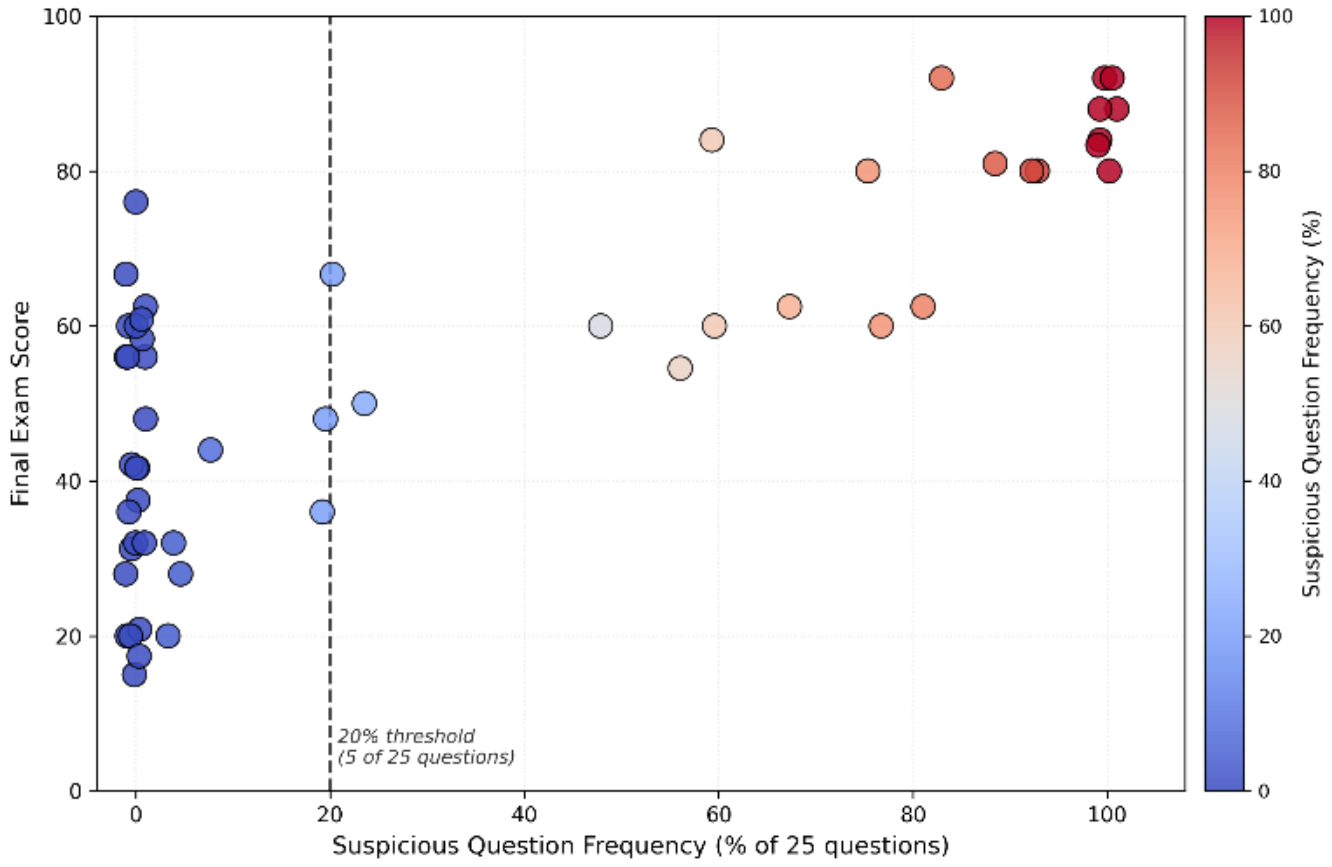

Fig. 2. Final exam score versus suspicious question frequency for all 52 students

The distribution is strongly bimodal. Twenty-five students showed no suspicious behavior, and four showed it in only one or two questions. Most of the remaining students displayed suspicious behavior in more than half of the exam. No student falls between three and four suspicious questions, leaving an empty region between the groups. The 20% labeling threshold, corresponding to five of the 25 questions, was placed above this gap so that isolated events that may have benign explanations were not treated as systematic behavior. Because students had been explicitly instructed not to perform these actions, their repeated occurrence across several questions was treated as a risk indicator. This rule produced 23 high-risk and 29 low-risk students. Because the cut-off sits above an empty region, the labels are not highly sensitive to a small change in the threshold. Any cut-off between 9% and 20% produces the same partition, while thresholds up to 48% affect only the four observations closest to the boundary. A sensitivity check at 10%, 20%, and 30% preserved the overall group pattern.

Fig. 2 also shows that several students in the high-risk group achieved relatively high final-exam scores while displaying suspicious behavior across multiple questions.

### B. *Learning Management System Data*

Learning activity was recorded using standard Moodle logs and a custom video player developed by the author. The player recorded video interactions, including playback, pausing, seeking, and watch time. When opening a video, students selected their viewing purpose, which was also recorded. Only data from the first eight weeks of the semester, before the midterm exam, were used. For each student, 27 features were derived in three categories: (i) video interaction features (n = 9); (ii) viewing-purpose features (n = 5), based on students' selections of reviewing after class, preparing before class, studying for exams, completing assignments, or other purposes; and (iii) Moodle activity features (n = 13), including sessions, total activity, course-module views, assignment views, assignment submissions, resource views, quiz attempts, and URL clicks. Exam, assignment, and quiz grades were excluded from the predictors.

### C. *Data Pre-Processing and Feature Selection*

Important scale differences were observed among the features in the dataset. MinMax scaling was therefore applied to Logistic Regression within each training fold because the model is sensitive to feature scale. Since tree-based algorithms are scale-independent, normalization was not applied in these models.

Feature selection was used to reduce dimensionality and noise. Within each Leave-One-Out training fold, the discriminating power of the 27 candidate features was assessed with the non-parametric Mann-Whitney U test, and the five features with the smallest p-values were retained for that fold. The number of retained features was fixed at five to constrain model complexity under small-sample conditions and to keep the resulting model interpretable. The held-out student data was not used in feature selection.

### D. *Classification Models and Evaluation*

The prediction of students' AI-assisted cheating risk was formulated as a binary classification problem. Four algorithms were compared: (1) Logistic Regression with MinMax scaling, (2) Gaussian Naive Bayes, (3) Random Forest with 200 trees, max_depth = 4, and class weighting, and (4) Gradient Boosting with 100 trees and max_depth = 2. These classifiers represent linear, probabilistic, bagging-based, and boosting-based approaches. Parameters not specified above were left at their scikit-learn defaults.

Given the small sample size (N = 52), model performance was evaluated using leave-one-out cross-validation (LOOCV) with fold-specific preprocessing and feature selection. In each of the 52 folds, one student was held out for testing, while the remaining 51 students formed the training set. All data-dependent steps, including scaling where required and feature selection using the Mann–Whitney U test, were performed using only the training set. The five selected features were then used to train the classifier, which was evaluated on the held-out student. Repeating this procedure produced one out-of-sample prediction for each student. Models were evaluated using Accuracy, Balanced Accuracy, Macro-F1, and the area under the ROC curve (AUC). Balanced Accuracy and AUC were treated as the primary metrics because the class sizes were unequal. All analyses were conducted with the Python scikit-learn library.

## III. RESULTS

### A. *Model Performance Comparison*

Table I summarizes the out-of-sample performance of the four classification models. The majority-class accuracy was 0.558 (29/52), and the chance level for AUC was 0.500; every model exceeded the majority-class accuracy baseline. Logistic Regression achieved the highest AUC and Balanced Accuracy (0.763 and 0.718, respectively). Naive Bayes produced a similar AUC of 0.760, whereas Random Forest and Gradient Boosting performed less well. For Logistic Regression and Naive Bayes, Balanced Accuracy remained close to raw Accuracy (0.718 versus 0.731 and 0.701 versus 0.712, respectively).

TABLE I. CLASSIFICATION PERFORMANCE UNDER LEAVE-ONE-OUT CROSS-VALIDATION WITH FOLD-SPECIFIC FEATURE SELECTION

| Model | Accuracy | Bal. Accuracy | Macro-F1 | AUC |
|---|---|---|---|---|
| Logistic Regression | 0.731 | 0.718 | 0.720 | 0.763 |
| Naive Bayes | 0.712 | 0.701 | 0.703 | 0.760 |
| Random Forest | 0.654 | 0.645 | 0.645 | 0.720 |
| Gradient Boosting | 0.673 | 0.666 | 0.667 | 0.646 |

This indicates that the mild class imbalance did not dominate the model ranking. The next analyses therefore examine the held-out predictions from Logistic Regression and the stability of the features selected across folds.

### B. *Detailed Analysis of the Best Model*

The held-out predictions from Logistic Regression were examined using a confusion matrix and an ROC curve. Fig. 3 presents both summaries.

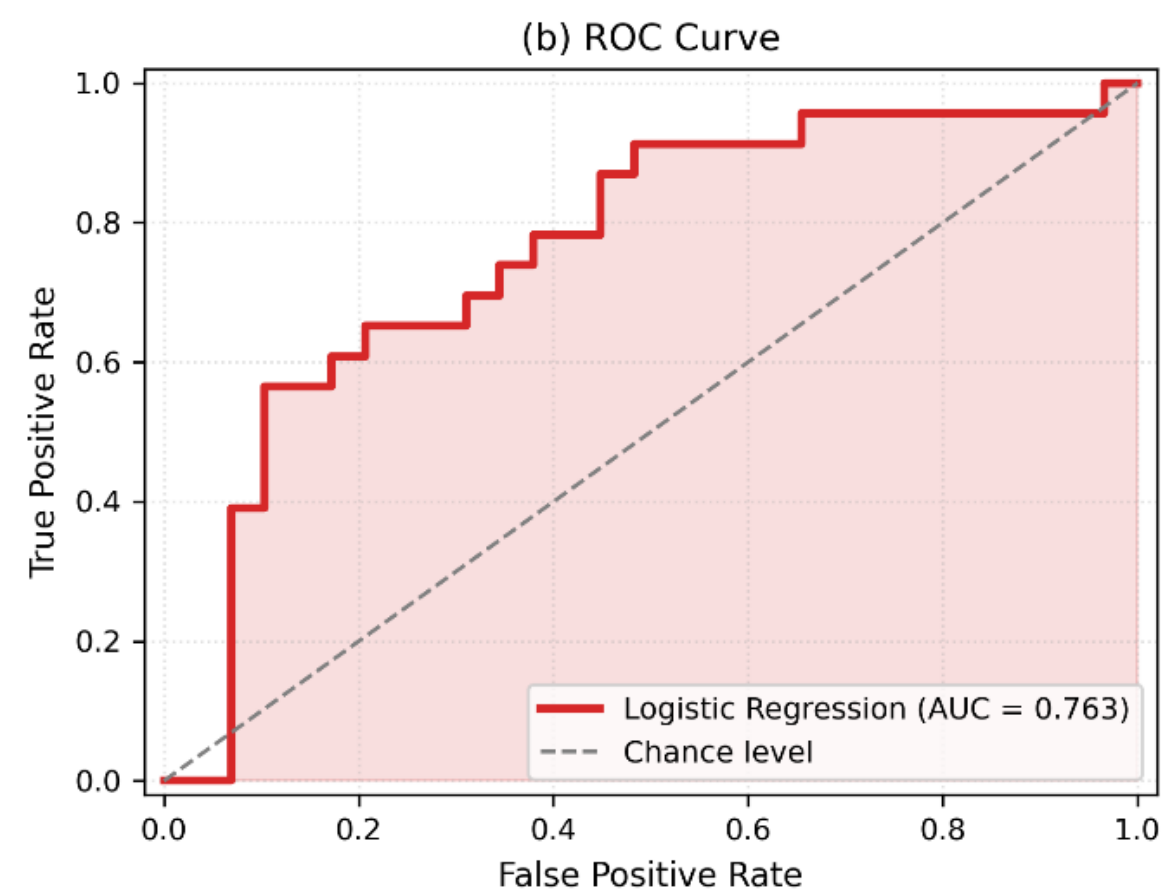

Fig. 3. Confusion matrix and ROC curve for Logistic Regression under LOOCV

Logistic Regression correctly identified 14 of the 23 high-risk students (Sensitivity = 60.9%) and 24 of the 29 low-risk students (Specificity = 82.8%). Among the 19 students predicted as high-risk, 14 belonged to the high-risk class (Precision = 73.7%). The AUC was 0.763.

These results use a classification decision threshold of 0.50. A lower threshold increases sensitivity but also increases false positives. At 0.30, the model identified 21 of 23 high-risk students (Sensitivity = 91.3%, Specificity = 51.7%, Precision = 60.0%). At 0.60, it identified 12 high-risk students (Sensitivity = 52.2%, Specificity = 89.7%, Precision = 80.0%). These operating points show why the threshold must be chosen according to the intended use of the prediction.

### C. *Feature Selection Stability and Interpretation*

For interpretation of the best-performing Logistic Regression pipeline, feature stability was assessed by counting how often each variable was selected across the 52 training folds (Table II). Three features were selected in every fold, one in 51 folds, and another in 45 folds. Only seven of the 27 candidates were selected in any fold. The selected feature set varied little across the 52 training folds.

Low-risk students had higher group means for all seven selected variables, with Cliff's delta values from 0.50 to 0.60. Selection frequencies describe the stability of the feature-screening procedure, while group means and Cliff's delta summarize the direction and magnitude of between-group differences. These summaries do not quantify model-specific feature importance. Overall, course-module views, assignment submissions, and video-access days were the most consistently selected features across the LOOCV folds, and all three had lower values in the high-risk group than in the low-risk group.

TABLE II. FOLD-WISE FEATURE SELECTION FREQUENCY, GROUP DESCRIPTIVES, AND CLIFF'S DELTA EFFECT SIZES

| Feature | Folds selected | Mean (low-risk) | Mean (high-risk) | Cliff's δ |
|---|---|---|---|---|
| LMS_ModuleViews | 52/52 (100%) | 177.41 | 96.26 | 0.60 |
| LMS_AssignmentSubmissions | 52/52 (100%) | 8.83 | 4.00 | 0.57 |
| VID_ActiveDays | 52/52 (100%) | 7.14 | 3.30 | 0.55 |
| VID_UniqueVideos | 51/52 (98%) | 7.34 | 4.00 | 0.54 |
| LMS_AssignmentViews | 45/52 (87%) | 99.52 | 48.04 | 0.53 |
| LMS_ResourceViews | 5/52 (10%) | 50.07 | 27.65 | 0.52 |
| LMS_LinkClicks | 3/52 (6%) | 23.93 | 13.78 | 0.50 |

## IV. DISCUSSION AND CONCLUSION

The study found that LMS traces from the first eight weeks could distinguish between students assigned to high-risk and low-risk groups in the final exam. Logistic Regression achieved an AUC of 0.763 and a Balanced Accuracy of 0.718. This result extends earlier learning-analytics research on using observable engagement data for early risk identification [12, 13] to an academic-integrity outcome. Course-module views, assignment submissions, and video-access measures were consistently lower in the high-risk group, indicating an association between lower recorded LMS engagement and higher behavioral risk. LMS traces, however, reflect observable activity rather than cognitive engagement [12]. The prediction should therefore be interpreted as a risk estimate, not as evidence of misconduct.

Fig. 2 provides a further descriptive check on the behavioral risk labels. Overall, 13 of the 23 high-risk students achieved a final-exam score of 80 or above, whereas none of the 29 low-risk students reached 80; the highest score in the low-risk group was 76. This contrast was particularly marked among the 12 students who displayed suspicious events in at least 20 of the 25 questions: 11 scored 80 or above, while the remaining student scored 62.5. Final-exam scores were excluded from label construction and model training. Their association with repeated suspicious events provides additional descriptive context, but it does not independently validate the risk labels. It is consistent with the possibility that unauthorized assistance contributed to the performance of some high-risk students.

The feature-selection results also offer a possible self-regulatory interpretation. Course-module views primarily reflect behavioral engagement, whereas the number of days on which course videos were accessed reflects how study activity was distributed over time, and assignment submissions reflect the fulfillment of course responsibilities. Previous learning-analytics research has used regularity, assignment submission, distributed access, and anti-procrastination behaviors as trace-based indicators of self-regulated learning [15]. The lower values observed in the high-risk group may therefore reflect weaker self-regulation as well as lower behavioral engagement. A review of academic cheating in online learning environments found that its relationship with self-regulation varies across contexts [16]. However, these features are trace-based proxies, and the present study did not directly measure self-regulated learning. Future studies should combine LMS traces with validated self-regulated learning measures to examine whether self-regulation predicts AI-assisted cheating risk or explains part of the association between early LMS engagement and subsequent risk.

Logistic Regression and Naive Bayes performed better than the two tree-based models in this dataset. These results suggest that relatively simple classifiers were competitive in this small dataset. The result is specific to the present dataset and should not be interpreted as general evidence that simple models will outperform more complex models in other courses or larger samples.

The practical value of the model depends on how its output is used. At a classification decision threshold of 0.50, specificity exceeded sensitivity. Lowering this threshold to 0.30 increased sensitivity to 91.3% but also increased false positives. A lower threshold may be suitable for low-stakes outreach, such as reminders, academic support, or guidance on acceptable AI use. It would not be suitable for disciplinary decisions. Instructors should review each flag in context, and the score should remain separate from grading and misconduct records.

This support-oriented interpretation is consistent with recent calls to respond to AI-enabled misconduct through assessment design and ethical pedagogy rather than relying only on surveillance [7]. The selected features point to concrete forms of support, including reminders about course materials, assignment follow-up, and brief instructor check-ins. Clear guidance is also important because students do not always interpret the boundary between acceptable AI assistance and cheating in the same way [6]. Any use of the model should therefore include human review, transparent communication, and a means for students to question how their data were interpreted.

Several limitations constrain the interpretation of these findings. The sample is small and comes from one programming course at one public university in Türkiye. Engagement patterns may differ with course design, LMS configuration, exam format, institutional AI policies, and local patterns of technology access. Systematic reviews also caution that LMS measures are context-dependent proxies for engagement [12, 13]. Both the risk-labeling rule and the classification decision threshold require local validation before use in another setting. The risk labels were derived from repeated prohibited events using a 20% labeling threshold, corresponding to five of the 25 questions. Their association with final-exam scores provides convergent but indirect support, not confirmation of cheating. Future studies should validate the labels through ethically collected evidence such as expert review, student interviews, or additional session records. Larger multi-course and multi-institutional studies should also examine temporal features and test whether supportive interventions improve outcomes. The present model is an early demonstration, not a deployment-ready system.

In conclusion, LMS interactions recorded during the first eight weeks of the semester may provide an early signal of later AI-assisted cheating risk. Moreover, 44.2% of students were labeled as high-risk in a proctored, face-to-face exam, which suggests that the issue warrants serious attention. The main contribution is the shift from detecting suspicious behavior during or after an exam to identifying students who may benefit from support before the exam. When used with human review and limited to low-stakes interventions, this approach may help institutions protect academic integrity without treating a prediction as an accusation. Replication with larger and more diverse samples, together with stronger validation of the risk labels, is required before operational use.

## ACKNOWLEDGMENT

The author used ChatGPT (OpenAI) for translation and to improve the language and readability of the manuscript. The author reviewed and edited the text as needed and takes full responsibility for the content of the publication.